# Tourism networks and computer networks

Rodolfo Baggio

rodolfo.baggio@unibocconi.it

Master in Economics and Tourism, Bocconi University, Milan, Italy

and

School of Tourism, The University of Queensland, Australia

Jan 2008

**Abstract**

The body of knowledge accumulated in recent years on the structure and the dynamics of complex networks has offered useful insights on the behaviour of many natural and artificial complex systems. The analysis of some of these, namely those formed by companies and institutions, however, has proved problematical mainly for the difficulties in collecting a reasonable amount of data. This contribution argues that the World Wide Web can provide an efficient and effective way to gather significant samples of networked socio-economic systems to be used for network analyses and simulations. The case discussed refers to a tourism destination, the fundamental subsystem of an industry which can be considered one of the most important in today's World economy.

**Introduction**

The title of this contribution mimics the one written by Wellman [1] some years ago. In that one, the author was suggesting a close relationship between a computer network and the social network of the people using it. In his words: *computer networks are inherently social networks, linking people, organizations, and knowledge.* Beside this, quite a number of studies have, in one way or another, maintained a similar interpretation.

This letter aims at examining the relationships between the technological World Wide Web (WWW) network and the complex socio-economic network formed by the companies and organisations grouped in a tourism destination.

In the second part of last century, tourism has become probably the largest economic sector of the World economy. The boundaries of the tourism and travel industry are fairly indefinite. It comprises a wide variety of organisations offering diverse products and services and exhibiting very little homogeneity. A tourism destination (TD), the place towards travellers head for to spend their time, can be broadly defined as a geographical area that offers the tourist the opportunity of exploiting a variety of attractions and services [2]. Scholars and practitioners consider it a fundamental unit of analysis for the understanding of the whole tourism sector. Essentially, a TD is a complex socio-economic system, the archetype of a *complex adaptive system* (CAS). It shares many (if not all) of the characteristics usually associated with a CAS: non-linear relationships among the components (private and public companies and associations), self-organisation and emergence of organisational structures, robustness to external shocks. A TD is based on a dynamic set of relationships, therefore a *network approach* would seem indispensable. In recent times, several authors have studied TDs using the perspective of chaos and complexity complex systems science [3, 4, 5], and a number of characteristics has been visibly identified both from a qualitative and a quantitative viewpoint. Only a few, however, have applied the methods and tools of the "network science" to improve our knowledge of the structure and the dynamical behaviour of a tourism system .

The Internet age has produced a wealth of new ways for producing and distributing travel and tourism services. Web-based approaches and technologies are helping suppliers and agencies in reducing service costs and attracting customers [6]. A website looks to be a major mechanism (and someone maintains it will be the only one in the future) to conduct business in the tourism field.

In many other fields, the Internet and the WWW network have given the basic materials with which network scientists have significantly improved our comprehension of how systems of any kind, not only computerised, are structured or behave [7]. The statistical mechanics tradition, then, has provided a strong theoretical framework in which these works can be embedded. Today we are ever more convinced that the properties of a complex network are not simply an *interesting curiosity*. They are bound to the intrinsic characteristics of the systems they represent [8, 9]. The analysis of hundreds of different networks and of their properties has allowed the formulation of quite a number of theories and models which have proved very effective also form a "practical" point of view. In this scenario, probably the least investigated are the social and economic networks.

The discipline known as *social network analysis* (SNA), has for many years collected results on the formation and the evolution of human, social and economic relationships, on the importance of some positions in the web of connections we have, and on how to use these outcomes to steer and to encourage the development of a community, a company, a society [10].

One major problem faced by SNA has always been the collection of the data needed for the analyses. Many methods have been devised and many techniques have been proposed to allow extracting meaningful insights from the sometimes scarce records a researcher is able to collect on the elements and the linkages of a social network [11].

On the other hand, the outcomes obtained by the contemporary network scientists could be used to improve this work. But these are typically coming from large quantities of data and at least a "decent" and reliable amount of it is needed to highlight structures, differences, patterns and to trace evolutionary processes and developments [see for example Refs. 12, 13].

**Real and virtual tourism networks**

The Web has been seen as a good candidate to provide this amount of data and a several studies have followed this direction creating a discipline called *hyperlink analysis* [14]. Some have argued and warned of the risks and the dangers of this approach, claiming that the links are created in a rather unpredictable way, and it is not possible to find unambiguous meanings [15]. While this can be true when thinking to webpages built by individuals, the situation looks different for private or public organisations. In many cases the practice of hyperlinking is regulated, and the presence of a link reflects a specific choice made by the website owner. A link is considered to be a strategic resource, and the possible variations in the structure of the "corporate" WWW are shaped by specific communicative aims, rather than by random technological processes [16].

A scrutiny of corporate websites (140 of them have been inspected) shows that a vast majority (87%) publishes some form of "instructions" and legal disclaimers about the website and its contents. All of them specifically address the issue of linking. They state how and when permissions must be sought before a connection can be made to the organisation's Web site. In addition, the analysis of a number of internal documents shows that carefully defined policies regulate the publication process for all the contents of a website, including the hyperlinks to other organisations [17].

Moreover, in a series of studies conducted in Italy, more than 400 tourism operators, grouped in a number of associations or consortia, have been informally asked to describe the process in place to update their websites. All of them, even those relying on some form of technical outsourcing, have stated that the insertion of a link towards a different *entity* is decided largely on a *business* basis. More specifically, links to companies belonging to the same sector of activity (and in the same geographic area) are only present when some kind of collaboration agreement exist [17].

A more rigorous study on the web networks of two tourism destinations, the island of Elba (Italy) and the Fiji islands [12], has shown that 68% of their links connect other tourism companies in the same area, 3% non-tourism companies in the same area, 22% tourism organisations outside the area, and the remaining 7% are generic "broad interest" links.

These outcomes, however, could be classified as *presumptive evidence*. A better way to specify the relationship between the virtual and the real network of a socio-economic system (a TD in the present instance) is to look for similarities and differences in their topological structures.

**Topological comparison of real and virtual networks**

The case presented here concerns the island of Elba, a well known Italian destination. Located in the centre of the Tyrrhenian sea, the island is a typical "sun and sand" destination. Almost all of its economy depends on the wealth generated by almost 500 000 tourists visiting the island every year.

The "real" connections among the tourism system stakeholders (TN), hotels, tourism agencies, service companies, restaurants, associations, have been enumerated by consulting publicly available documents such as membership lists for associations and consortia, commercial publications, ownership and board of directors records. The data obtained and the completeness have been validated with a series of structured and unstructured interviews to a selected sample of local *knowledgeable informants* such as the directors of the local tourism board and of the main industrial associations. The web network (WN) has been collected by listing all the hyperlinks among the websites belonging to the tourism operators located on the island [13].

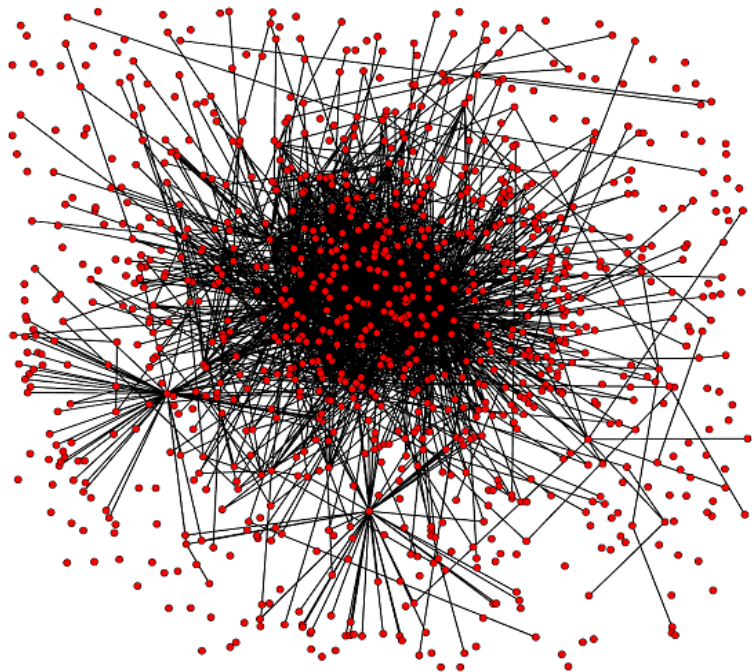

Fig. 1 Elba tourism network, the connections among the tourism organisations located in the island of Elba, Italy

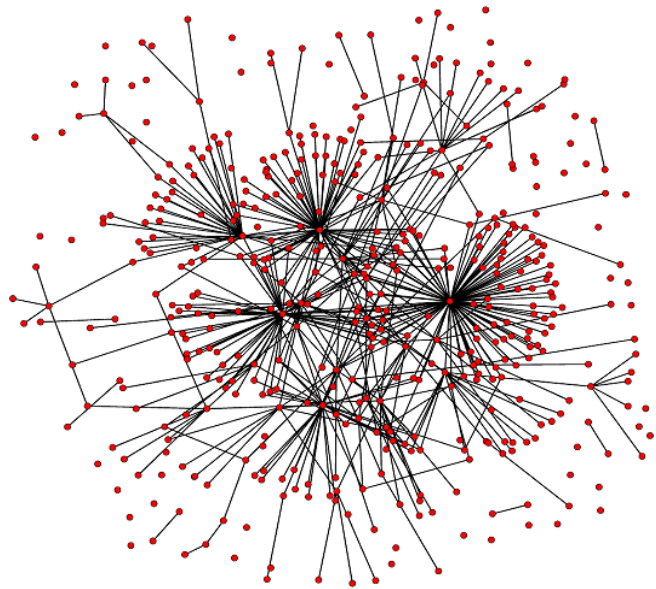
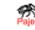

Fig. 2 Elba tourism web network. Edges are the hyperlinks among the websites belonging to the tourism organisations located in the island of Elba, Italy

The two networks (Fig. 1 and Fig. 2) have been then investigated by using standard network analysis techniques. Table 1 shows the values for the main metrics calculated for them. As it can be seen, apart from scale factors, most of the values have differences which are lower than an order of magnitude.

One more important indicator is the degree distribution which is commonly regarded as a signature of the network topology. The cumulative degree distributions are shown in fig. 3. The exponents of the power-law degree distribution calculated from this data are: TN = 2.32±0.269; WN: 2.19±0.109 [calculations have been performed according to Ref. 18]. They can be considered identical within the statistical uncertainty of their determination.

It is known from the literature [see for example Refs. 19, 20] that in most cases the various quantities characterising the topology of a complex network can hardly be considered normally distributed, and the simple comparison of their averages may look insufficient.

Table 1 Main network metrics and characteristics of the *TN* and *WN* networks. The values have been obtained by using available software packages (Pajek, Ucinet) complemented by some Matlab programs developed by the author. Degree distribution scaling exponents are calculated according to Ref. [18].

| Metric | TN | WN |
|---|---:|---:|
| Number of nodes | 1028 | 468 |
| Number of edges | 1642 | 495 |
| Density | 0.003 | 0.005 |
| Disconnected nodes | 37% | 21% |
| Diameter | 8 | 10 |
| Average path length | 3.16 | 3.70 |
| Clustering coefficient | 0.050 | 0.014 |
| Degree distribution exponent | 2.32 | 2.17 |
| Proximity ratio | 34.10 | 12.21 |
| Average degree | 3.19 | 2.12 |
| Average closeness | 0.121 | 0.155 |
| Average betweenness | 0.001 | 0.003 |
| Global efficiency | 0.131 | 0.170 |
| Local efficiency | 0.062 | 0.015 |
| Assortativity coefficient | -0.164 | -0.167 |

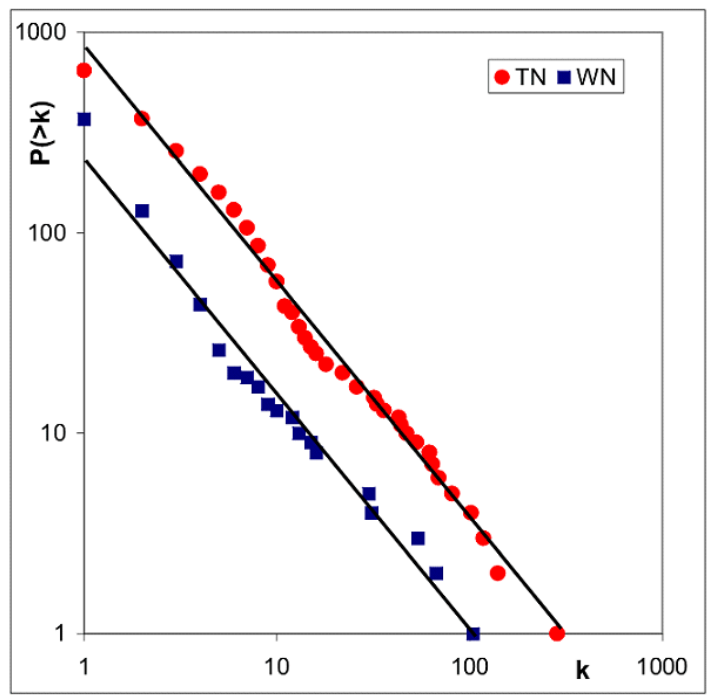

Fig. 3 Cumulative degree distributions for the tourism networks examined. WN is the network of websites belonging to the tourism stakeholders in the island of Elba, TN is the one formed by considering the "real" linkages among them.

In these cases, as already proposed by some researchers [18, 21], the Kolmogorov-Smirnov (KS) statistic is able to provide good results. The KS D-statistic gives the maximum distance between the cumulative probability distributions of empirical data F(x) and G(x) over the entire x range: $D = \max_x |F(x) - G(x)|$. The statistic is nonparametric and it is insensitive to scaling issues, it compares only the shapes of the empirical distributions [22].

Table 2 shows the values for the D-statistics calculated when comparing the quantities of the Web network with those of the real network (WN vs. TN). As reference, the same values have been calculated for a random sample of the same size as WN, extracted from the real one (RN vs. TN: the values are averages over ten realisations). The consistently lower values of the D-statistic in the case of the web network can be considered as a good confirmation of the similarities of the two topologies.

Table 2 The Kolmogorov-Smirnov D-statistics. Values obtained by relating the distributions of the different network metrics. The *WN vs. TN* column refers to the comparison between the Web network and the real network, *RN vs. TN* considers the real network and a sample obtained by randomly choosing a number of nodes equal to the one composing the WN network (values are averages over ten realisations).

| Metric | WN vs. TN | RN vs. TN |
|---|---|---|
| Degrees | 0.119 | 0.147 |
| Clustering coefficient | 0.147 | 0.178 |
| Closeness | 0.044 | 0.083 |
| Betweenness | 0.030 | 0.077 |
| Local efficiency | 0.125 | 0.184 |

**Concluding remarks**

A number of considerations and a quantitative study of a real and a virtual set of linkages among the tourism operators located in a destination has shown a strong similarity between the two networks.

Clearly, a full equality cannot be claimed, but the results reported here can legitimate a researcher in using the Web as the source to collect a significant sample of the underlying socio-economic network.

The obvious limitation is that the comparison holds when considering "institutional" websites, belonging to companies, associations or other institutions. Moreover, the area taken into account must show a quite high diffusion of the Internet and the Web. Yet nowadays this one, for a large part of the World, is not a severe limitation.

By carefully applying the considerations made in the literature [23, 24] on the handling of network samples, the WWW can proof again an effective environment to study the characteristics and the behaviour of social and economic systems formed by companies, corporations, associations and other such entities. Tools and methods of the science of networks can thus be extended to these important elements in today's World.

The importance of the results presented here is even higher for the field of tourism. As stated at the beginning of this letter, a tourism destination is probably the most important component of this system. Many recent quantitative network analysis methods can provide more possibilities to improve and complement our knowledge of these structures. Moreover, the network approach can be extended to implement simulation models with which different scenarios can be obtained in order to explore the possible effects of different managerial activities. This would give all people interested in the life of a tourism destination powerful tools to inform their policy or management actions.